\documentclass[%
 aip,
 amsmath,amssymb,
 reprint,%
]{revtex4-1}

\usepackage{graphicx}
\usepackage{dcolumn}
\usepackage{bm}

\usepackage[utf8]{inputenc}
\usepackage[T1]{fontenc}
\usepackage{mathptmx}
\usepackage{color}
\usepackage[colorlinks=true,bookmarks=false,citecolor=blue,urlcolor=blue]{hyperref}

\begin{document}

\preprint{AIP/123-QED}

\title[Tutorial]{Tutorial: synthetic frequency dimensions in dynamically modulated ring resonators}

\author{Luqi Yuan}
\email{yuanluqi@sjtu.edu.cn}
\affiliation{State Key Laboratory of Advanced Optical
Communication Systems and Networks,
School of Physics and Astronomy, Shanghai Jiao Tong University, Shanghai 200240, China
}%

\author{Avik Dutt}

\author{Shanhui Fan}
\email{shanhui@stanford.edu}
\affiliation{%
Ginzton Laboratory and Department of Electrical Engineering,
Stanford University, Stanford, CA 94305, USA
}%

\date{\today}

\begin{abstract}
The concept of synthetic dimensions in photonics has attracted rapidly growing interest in the past few years. Among a variety of photonic systems, the ring resonator system under dynamic modulation has been investigated in depth both in theory and experiment, and has proven to be a powerful way to build synthetic frequency dimensions. In this tutorial, we start with a pedagogical introduction to the theoretical approaches in describing the dynamically modulated ring resonator system, and then review experimental methods in building such a system. Moreover, we discuss important physical phenomena in synthetic dimensions, including nontrivial topological physics. Our tutorial provides a pathway towards studying the dynamically modulated ring resonator system, understanding synthetic dimensions in photonics, and discusses future prospects for both fundamental research and practical applications using synthetic dimensions.
\end{abstract}

\maketitle

\section{Introduction}

In physics, dimensionality, i.e. the number of independent directions in a system, is a key factor for characterizing a
broad range of physical phenomena, affecting a variety of physical
dynamics \cite{feymannbook}.  Physical phenomena can have very
different characteristics depending on the dimensions of the physical systems. For
example, both Anderson localization and recurrence properties of random walks have different characteristics in one or two dimensions as compared to three or more dimensions \cite{anderson58,lee85,lagendijk09}. Another example is the spontaneous symmetry-breaking phase transition in the ferromagnetic Ising model, which does not occur in one dimension (1D) but does occur in two or more dimensions \cite{brush67,elshowk12,cosme15,belamotte16}.  The explicit dimensions in the physical world include spatial dimensions, which could be up to three dimensions at most, as well as a temporal dimension, although extra space-time dimensions have been theorized in high energy physics \cite{rs1999prl,shifman10}. On the other hand, higher dimensional physics
such as four-, five- and six-dimensional topological physics has been
attracting great interest recently in theory \cite{zhang01,qi08,lian16,lian17, petrides18prb}, supporting exotic states of light and matter with no lower-dimensional analogues \cite{zhang01}. Moreover, lower-dimensional structures 
(such as those in 1D or 2D) are usually easier to build than structures in 
3D. Hence, significant efforts have focused on developing the concept of synthetic dimensions, which uses degrees of freedom of the physical system other
than spatial dimensions to augment the spatial dimensions in order to explore higher dimensional physics \cite{yuan18optica,ozawa19nrp}.

In the past decade, synthetic dimensions have been extensively
explored in systems including cold atoms \cite{boada12,mei12,celi14,price15,mancini15,stuhl15,wang15,martin17,baum18,lohse18,chen18,an18,zhang19,boyers20}, hot
atomic gases \cite{cai19,he21},  photonic platforms \cite{regensburger12, jukic13, schwartz13, luo15, yuan16, ozawa16, lustig19, dutt20}, superconducting circuits~\cite{tsomokos2010pra, lee20pra, hung_quantum_2021}, and optomechanics~\cite{schmidt2015optica}. 
In the photonic implementations, for the quantum (few-photon) regime,  synthetic lattices  with arbitrary dimensions were constructed using the Fock-state ladder formed by an atom coupled to several multimode cavities \cite{wang16,cai21}. For classical light, several internal degrees of freedom can be utilized to synthesize new dimensions \cite{yuan18optica}. Two prominent methods 
to introduce synthetic dimensions in photonics are (i) to build connectivity between different states of the system \cite{regensburger12,jukic13,schwartz13,luo15,yuan16,ozawa16,lustig19,dutt20},
and (ii) to use a tunable parameter in the system and to consider the parameter dimensions as the extra dimensions \cite{naumis08,kraus12,verbin13,wang17,zilberberg18,hu20CP,wang20OE}. The former method is an efficient approach to construct
synthetic lattices with complicated hopping coefficients, and
therefore provides important platforms to not only simulate interesting higher
dimensional \textit{states} but also study exotic \textit{dynamics}
different from those in spatial dimensions \cite{bell17,qin18prl,yuan18prb,lin18,yu20}. The latter approach, although limited in its ability to study dynamics, has shown promise in experimentally emulating 4D quantum Hall Hamiltonians directly in reciprocal space~\cite{kraus12,zilberberg18}. So far, various photonic degrees of freedom have been proposed to construct synthetic dimensions using the former method, including
the arrival time of pulses \cite{regensburger12}, the orbital angular momentum of light
\cite{luo15}, the frequency of the photon \cite{yuan16,ozawa16}, and the spatial supermodes of waveguides \cite{lustig19}. Photonic synthetic dimensions not only lead to the demonstrations of a variety of fundamental physics effects, but also triggers explorations towards potential applications in optical communications and quantum simulation \cite{yuan18optica,ozawa19nrp}.

In this paper, we provide a tutorial on  the creation of synthetic frequency
dimensions in the dynamically modulated ring resonator system. We focus on this system since there have been significant developments in experimental implementation of synthetic dimensions using this system. Many of the concepts developed in this system are generally applicable for other synthetic dimension systems as well. 
Theoretical approaches are outlined in Section II.  In
Section III, we introduce experimental methods to construct the system and to measure its band structure. We overview
examples of physics of synthetic dimensions as studied in this system  in Section IV.
We finish our article by providing a brief outlook  and a summary in Section V. We hope that
our work gives an illustrative picture of constructing synthetic
dimensions with engineered connectivity along the frequency axis of light and can trigger  more interesting ideas associated
with synthetic dimensions in photonics.

\section{Theoretical approaches}

\begin{figure}[htbp]
 \centering
 \includegraphics[width=0.48\textwidth ]{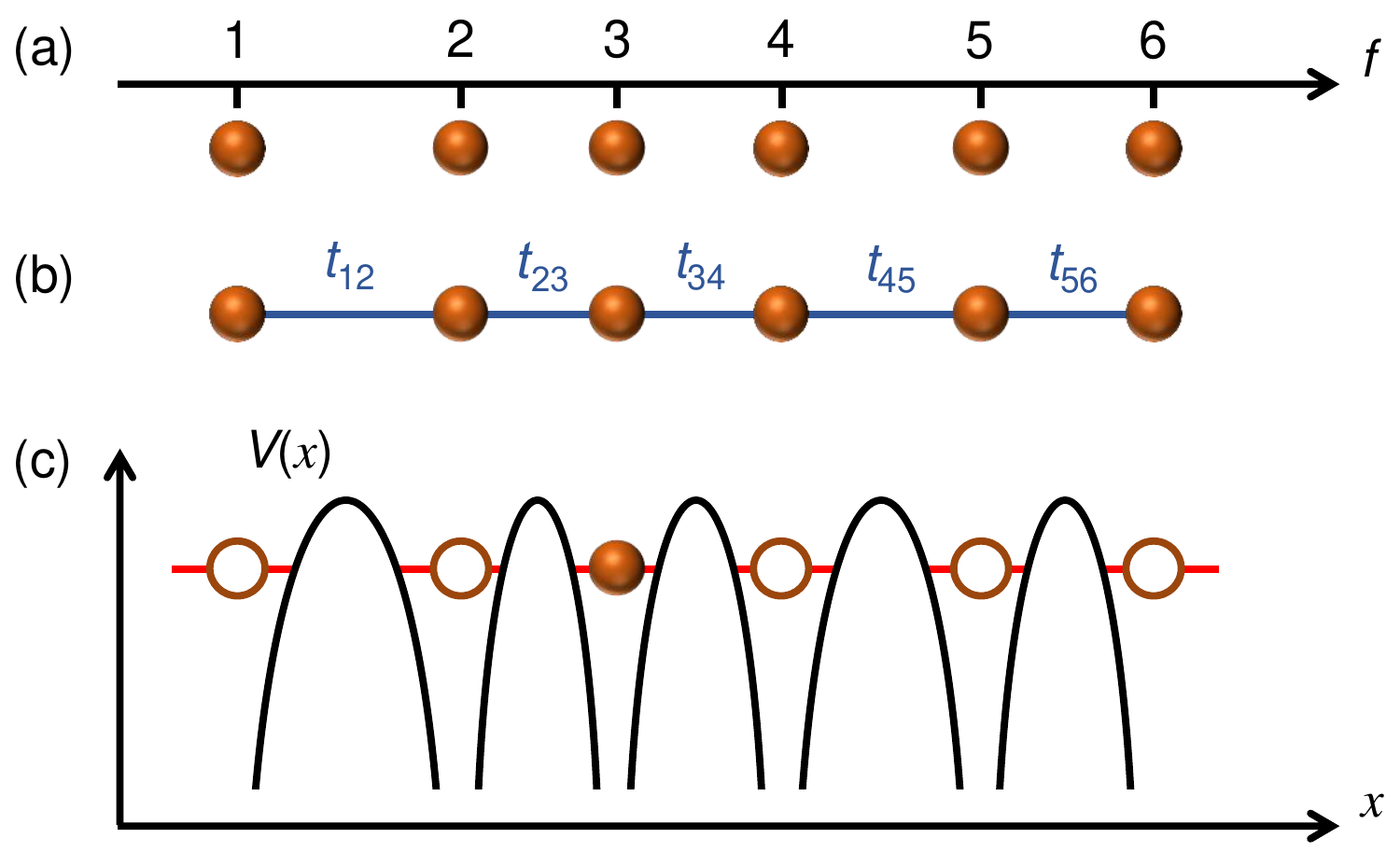}
 \caption{(a) Discrete resonant modes at different frequencies. (b) Modes are connected in the sequential order to build the synthetic dimension. (c) Schematic of an atom trapped at various potential wells. The dynamics of the atom can be described by the tight binding approximation. }\label{figure.1}
\end{figure}

The essential ingredient for creating a synthetic dimension is to build artificial connectivity between different physical states and connect them in a particular order, enabling the system to mimic extra spatial dimensions. Take the frequency axis of light as an example: the frequency axis is a \textit{dimension}; yet, as one learns from any textbook, two beams of light prepared in different frequencies can not transfer energy between each other in a static linear system. Hence, the frequency dimension in nature lacks connectivity to simulate the physics in the real space. However, if one considers discrete optical modes at different frequencies and one also finds a way to connect these discrete modes, the energy in modes at different frequencies can then be transferred between them, or in other words, the energy in one optical mode can hop into nearby modes along the frequency axis of light. Such artificial connectivity between optical modes at different frequencies may then be used to mimic the dynamics of one particle evolving in a one-dimensional lattice composed of multiple potential wells along the spatial dimension (see Fig. 1).  The example above (as shown in Fig. \ref{figure.1}) then gives a one-dimensional \textit{synthetic frequency dimension}. 

In this section, we briefly summarize theoretical methods to describe the constructed synthetic frequency dimension in the dynamically modulated ring resonator system, including the tight-binding model (and coupled-mode equations), the scattering matrix approach, and simulations with wave equations.

\subsection{The tight-binding model}

In photonics, for a variety of systems that have been used to construct synthetic dimensions using different degrees of freedom of light \cite{yuan18optica}, the tight-binding model has turned out to be a very powerful method to describe theoretically the physics of the systems. We again take the simple one-dimensional discrete physical modes shown in Fig. \ref{figure.1} that are artificially connected in a sequential order to construct the synthetic dimension as an example. It can be described by a Hamiltonian:
\begin{equation}
H_{\mathrm{TB}} = \sum_m V_m a_m^\dagger a_m + \sum_m \left( g_m a_{m+1}^\dagger a_m e^{i \phi_m} + h.c. \right), \label{Eq1:TB}
\end{equation}
where $a_m^\dagger$($a_m$) is the creation (annihilation) operator and $V_m$ is the effective onsite potential for the $m$-th mode. $g_m$ and $\phi_m$ give the hopping amplitude as well as the hopping phase between the $m$-th and $(m+1)$-th modes due to the artificial connectivity. $\hbar=1$ for simplicity. One can clearly see that the Hamiltonian in Eq. (\ref{Eq1:TB}) is the same as that describes the one-dimensional lattice composed of multiple potential wells in the tight-binding limit \cite{SSPbook}, so one creates the synthetic dimension following this way. 

The tight-binding model has been widely used to describe the synthetic dimension generated by utilizing different degrees of freedom of light, including the arrival time of pulses \cite{regensburger11}, the orbital angular momentum \cite{luo17,sun17,zhou17,luo18}, the frequency \cite{yuan16optica,lin16,zhang17,yuan17,yuan18APLP,qin18pra,qin18OE,li19,yang20,song20,dutt20light,yuan20,tusnin20}, and other degrees of freedom \cite{yu16,ozawa17,yuan19,zhang20}. In these systems the tight-binding model shows excellent capability of predicting and explaining experimental measurements. Here we focus  on the synthetic frequency dimension in a ring resonator under the dynamic modulation of the refractive index, which can be well described by the tight-binding model. 

\begin{figure}[htbp]
 \centering
 \includegraphics[width=0.48\textwidth ]{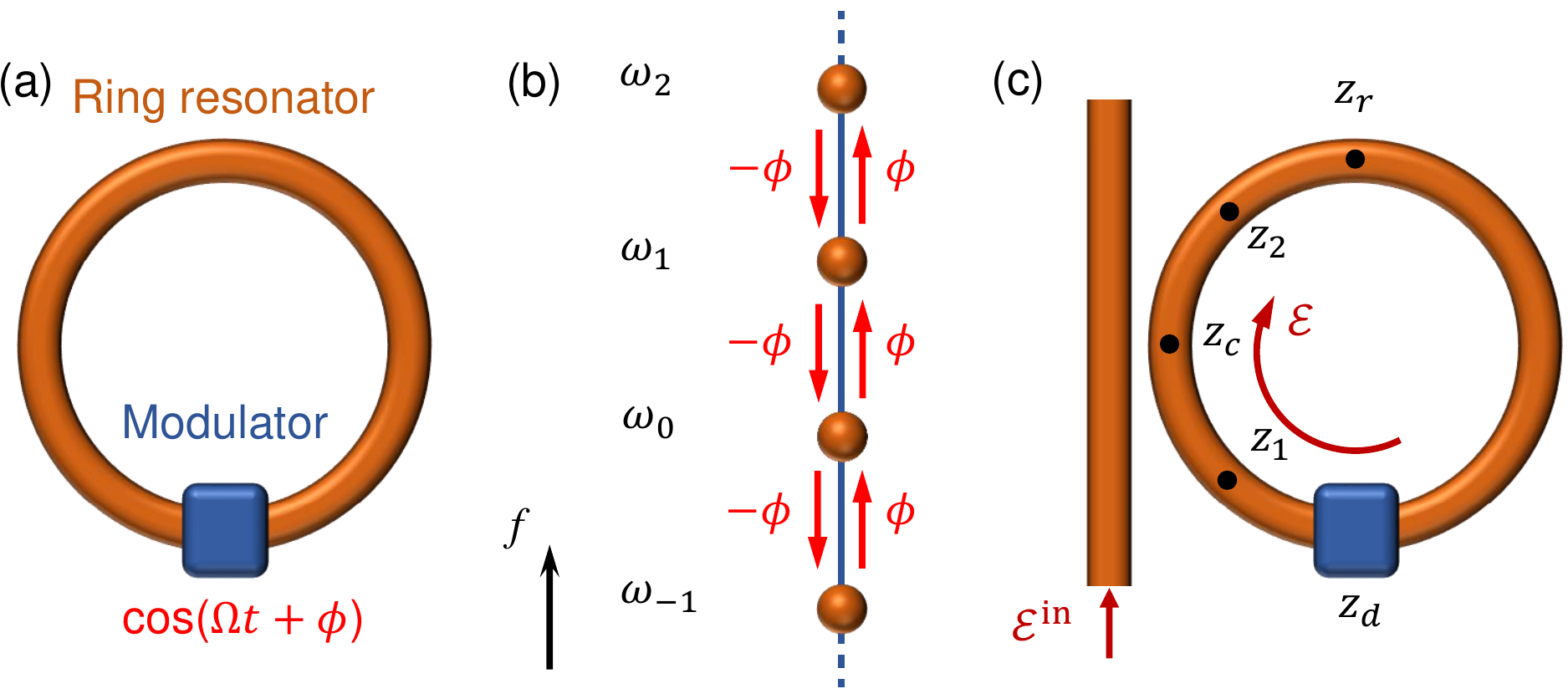}
 \caption{(a) A ring under the dynamic modulation. The ring in (a) supports a synthetic frequency dimension. (c) The modulated ring is coupled with external waveguide.}\label{figure.2}
\end{figure}

We consider a ring resonator made of a waveguide  with a length of $L$. Such a ring resonator  supports resonant modes with the frequency spacing $\Omega_R=2\pi v_g/ L$, where $v_g$ is the group velocity for light in the waveguide near the reference frequency $\omega_0$ (see Fig. \ref{figure.2}(a)). A modulator is placed inside the ring, with a time-dependent transmission coefficient \begin{equation}
T=e^{i2\kappa \cos (\Omega t +\phi)},  \label{Eq1:trans}
\end{equation}
where $\kappa$ is the modulation strength, $\Omega$ is the modulation frequency, and $\phi$ is the modulation phase. For the resonant modulation $\Omega=\Omega_R$, the corresponding tight-binding Hamiltonian is \cite{yuan16}
\begin{equation}
H_0 = g \sum_m  \left( a_{m+1}^\dagger a_m e^{i  \phi} + a_{m}^\dagger a_{m+1} e^{-i  \phi}  \right), \label{Eq1:TBring1}
\end{equation}
where $g=\kappa \Omega_R/2\pi$ is the hopping amplitude. Eq. (\ref{Eq1:TBring1}) shows that light at the $m$-th resonant mode (with frequency $\omega_m = \omega + m\Omega_R$) can be connected to its nearby modes, and thus a synthetic frequency dimension can be constructed through dynamic modulation.

\subsection{Effective gauge potential}

One notes that the modulation phase $\phi$ is imprinted into hopping coefficients in Eq. (\ref{Eq1:TBring1}) as shown in Fig. \ref{figure.2}(b). Importantly, the Hamiltonian in Eq. (\ref{Eq1:TBring1}) has the same form as the Hamiltonian of a charged particle in a lattice with a gauge potential \cite{luttinger51}. More specifically, if we use \cite{fang12}
\begin{equation}
\int_m^{m+1} A_{\mathrm{eff}}\  df = \phi \rightarrow A_{\mathrm{eff}}=\phi/\Omega_R, \label{Eq1:gaugeA}
\end{equation}
where the integration is along the frequency axis, we find that the modulation phase is linked to an effective gauge potential for photons in the synthetic frequency dimension. This leads to many interesting physical phenomena that we will discuss in Section IV.

\subsection{Theoretical opportunities with tight-binding Hamiltonians}

The tight-binding Hamiltonian description of the synthetic frequency dimension constructed in dynamically modulated ring resonators brings many theoretical opportunities. By adding another dimension \cite{yuan16,ozawa16} or folding the frequency dimension  \cite{yuan18prb} to construct a two-dimensional synthetic space, one can build non-uniform distribution of the effective gauge potential for photons and introduce the effective magnetic flux $B_{\mathrm{eff}}=\nabla \times A_{\mathrm{eff}}$, where $\nabla$ is the differential operator in the synthetic space.

Eq. (\ref{Eq1:TBring1}) gives the Hamiltonian of a ring resonator under the resonant modulation. For a more general case where the modulation is near resonance ($\Omega \approx \Omega_R$), the corresponding Hamiltonian is \cite{yuan16optica}
\begin{equation}
H = g \sum_m  \left( a_{m+1}^\dagger a_m e^{i \left(\Delta t + \phi\right)} + a_{m}^\dagger a_{m+1} e^{-i \left(\Delta t + \phi\right)} \right), \label{Eq1:TBring}
\end{equation}
where $\Delta = \Omega-\Omega_R$ is the offset between the modulation frequency and the frequency spacing between resonant modes.

The evolution of the system can be studied by assuming that the state of light has the form \cite{yuan16,yuan16optica}
\begin{equation}
|\Psi (t) \rangle = \sum_m C_m (t) a_m^\dagger |0\rangle, \label{Eq1:state}
\end{equation}
where $|0\rangle$ is the vacuum state and $C_m (t)$ denotes the amplitude of light at the $m$-th resonant mode. This form is valid to study the evolution of either the quantum dynamics of a single photon or the classical dynamics of coherent light. By plugging Eqs. (\ref{Eq1:TBring}) and (\ref{Eq1:state}) into the Schr\"{o}dinger equation $i d |\Psi (t) \rangle /dt = H |\Psi (t) \rangle $, one obtains
\begin{equation}
i \dot C_m (t) = g \left(C_{m-1} (t)e^{i \left(\Delta t + \phi\right)} + C_{m+1} (t)e^{-i \left(\Delta t + \phi\right)}  \right). \label{Eq1:CM}
\end{equation}
Eq. (\ref{Eq1:CM}) is the coupled-mode equation that can be used to simulate the dynamics of light in the dynamically-modulated ring.

It is interesting to note that the Hamiltonian in Eq. (\ref{Eq1:TBring}) is equivalent to a Hamiltonian describing a charged particle in a periodic lattice but subject to a constant electric field: \cite{yuan16optica}
\begin{equation}
H_2 =\sum_m m \Delta a_m^\dagger a_m + g \sum_m  \left( a_{m+1}^\dagger a_m e^{i \phi} + a_{m}^\dagger a_{m+1} e^{-i \phi} \right). \label{Eq1:TBforce}
\end{equation}
This can be proven using the gauge transformation \cite{yuan15,yuan16prb}
\begin{equation*}
|\Psi (t) \rangle = \sum_m C_m (t) a_m^\dagger |0\rangle \rightarrow 
\end{equation*}
\begin{equation}
|\tilde \Psi (t) \rangle = \sum_m \tilde C_m (t) a_m^\dagger |0\rangle = \sum_m  \left(C_m (t)e^{-im\Delta t}\right) a_m^\dagger |0\rangle, \label{Eq1:gaugeE}
\end{equation}
to show that $i d |\tilde \Psi (t) \rangle /dt = H_2 |\tilde\Psi (t) \rangle $ is satisfied. The first term on the right-side of Eq. (\ref{Eq1:TBforce}) corresponds to the effective potential $V_{\mathrm{eff}}=m \Delta$ linearly dependent on $m$ and then gives the effective electric field
$E_{\mathrm{eff}} =- \partial V_{\mathrm{eff}}/\partial f =- \Delta/\Omega_R$. 

Introducing multiple frequencies creates even more flexibilities for synthesizing Hamiltonians.  As an example, we consider the transmission coefficient
\begin{equation}
T=e^{i2\kappa \cos (\Omega t +\phi) + i2\kappa_N \cos (N\Omega t +\phi_N)},  \label{Eq1:trans2}
\end{equation}
which includes the lowest-order  resonant modulation $\Omega=\Omega_R$, as well as higher-order resonant modulations at frequencies $N\Omega$ with  strength $\kappa_N$ and phase $\phi_N$. Such a modulation leads to the tight-binding Hamiltonian 
\cite{yuan16}
\begin{equation}
H_3= g \sum_m  \left( a_{m+1}^\dagger a_m e^{i  \phi} + h.c. \right) + g_N \sum_m  \left( a_{m+N}^\dagger a_m e^{i  \phi_N} + h.c.  \right), \label{Eq1:TBlong}
\end{equation}
where $g_N=\kappa_N \Omega_R/2\pi$. One can see that this Hamiltonian includes not only the nearest-neighbor coupling, but also a long-range coupling between two lattice sites separated by a distance $N$. Therefore, a dynamically-modulated ring resonator system with multiple modulation frequencies can easily implement the long-range hopping process in the synthetic frequency dimension, which, on the other hand, can not be straightforwardly achieved in various real space implementation of such tight-binding model, such as through the use of coupled resonator array \cite{yariv99}. Such a long-range coupling may be used to achieve higher-dimensional space in a single modulated ring resonator \cite{schwartz13,yuan18prb}. 

In the setup of Fig. \ref{figure.2}(a), instead of using a phase modulator, one can use an amplitude modulator. In this case, 
the Hamiltonian to describe this system becomes \cite{yuan18APLP}
\begin{equation}
H_4 =  \sum_m  \left( h a_{m+1}^\dagger a_m e^{i  \phi_2} -h a_m^\dagger a_{m+1} e^{- i  \phi_2}. \right) , \label{Eq1:TBnH}
\end{equation}
where $h$ is the hopping amplitude and $\phi_2$ is the corresponding modulation phase. One notes that the coupling here is non-Hermitian. For a ring resonator that includes both a phase modulator and an amplitude modulator, the corresponding non-Hermitian Hamiltonian is \cite{kai21science}
\begin{equation}
H_5 =  \sum_m  \left\{(g + h) a_{m+1}^\dagger a_m e^{i(\phi + \phi_2)} + (g - h) a_m^\dagger a_{m+1} e^{-i(\phi + \phi_2)} \right\}. \label{Eq1:TBnH2}
\end{equation}
Here $g,h \in \mathcal{R}$. The ring resonator undergoing simultaneous phase and amplitude modulations is very useful for exploring non-Hermitian physics in the synthetic frequency dimension \cite{yuan18APLP,kai21science,longhi16}.

While any physical ring resonator can only support a finite number of resonances, theoretically it is very useful to consider a system where the number of modes along the frequency axis is infinite. For such an infinite system, the modal translational invariance in the frequency dimension gives rise to  the  momentum $k_f$ along the synthetic dimension. $k_f$ is a good quantum number for the infinite system. It is a conjugate variable  to frequency and therefore has a dimension of time, below we will show that indeed $k_f$ is a time variable. To illustrate how the $k_f$ arises,  we take the Fourier transform
\begin{equation}
a_{k_f} = \sum_m a_m e^{i m k_f \Omega_R}, \label{Eq1:FT}
\end{equation}
and the Hamiltonian in Eq. (\ref{Eq1:TBring}) becomes
\begin{equation}
H_{k_f} = 2g a_{k_f}^\dagger a_{k_f}\cos \left(k_f \Omega_R -\Delta t-\phi\right), \label{Eq1:Hk}
\end{equation}
In the case where $\Delta = 0$, the Hamiltonian in Eq. (\ref{Eq1:TBring}) is periodic along the frequency dimension and also time-independent, the eigenvalues of Eq. (\ref{Eq1:Hk}) then defines a bandstructure $\varepsilon= 2g \cos (k_f \Omega_R - \phi)$. When $\Delta  \neq 0$, the translational symmetry in Eq. (\ref{Eq1:TBring}) is still preserved. The eigenvalues of Eq. (\ref{Eq1:Hk}) then describes a dynamic bandstructure $\varepsilon (t)= 2g \cos (k_f \Omega_R \Delta t- \phi)$. Moreover, by comparing  Eq. (\ref{Eq1:FT}) with the total field in the ring, $E(t) = \sum_m C_m e^{i \omega_m t} = \sum_m C_m e^{i \omega_0 t + i m\Omega_R t}$, we notice  that $k_f$ is nothing but the time $t$. Due to the periodicity along the frequency dimension,  $k_f \in [0,2\pi/\Omega_R]$ in the first Brillouin zone physically corresponds to the time $t \in [0,T]$ where $T = 2\pi/\Omega_R = L/v_g$  is the round-trip time.

We end this section with several remarks. First, the Hamiltonian in Eq. (\ref{Eq1:TB}) is linear. But various nonlinear effects can be described by adding appropriate terms into the Hamiltonian. Second, in many experiments one probes the dynamics of light inside the ring by coupling it to external waveguides, as shown in Fig. \ref{figure.2}(c) for the case of one such external waveguide. The effect of coupling can be described using the input-output formalism \cite{gardiner85,fan10}, which describes the dynamics of the entire waveguide-resonator system in the Heisenberg picture. In the case with only one external waveguide as shown in Fig. \ref{figure.2}(c), the equations of the formalism reads:  \cite{yuan20}
\begin{equation}
\frac{d}{dt}a_m(t) = i [H_{\mathrm{TB}},a_m]- \frac{\gamma}{2} a_m(t) + i\sqrt{\gamma}\ c_{\mathrm{in},m}(t), \label{Eq1:Hh}
\end{equation}
\begin{equation}
c_{\mathrm{out},m}(t) = c_{\mathrm{in},m}(t) + i\sqrt{\gamma}\ a_{m}(t). \label{Eq1:cinout}
\end{equation}
Here $\gamma$ is the coupling strength between the system and the external waveguide, and $c_{\mathrm{in},m}$ ($c_{\mathrm{out},m}$) is the input (output) annihilation operator for the waveguide photons at the frequency of the $m$-th mode. The input-output formalism is a quantum formalism, but it is closely related to the temporal coupled mode theory formalism that describes the input/output process for the same system for the classical waves \cite{hausbook,fan03}. These formalisms are quite useful in the design of experiments, as we shall discuss in details in Section III.

\subsection{General wave equations}

Although the tight-binding model is perhaps the simplest approach in capturing the physics of a system exhibiting a synthetic dimension, it does not provide information associated with the spatial distributions of the modes.  We again take the ring resonator under the dynamic modulation as an example, the operator $a_m$ or the amplitude $C_m$ that we discussed in the previous section can describe the evolution of a single mode at the frequency $\omega_m$ at a certain reference position $z_r$ in the ring (see Fig. \ref{figure.2}(c)). Yet the information of the field propagating inside the ring at positions other than $z_r$ are not explicitly described in the tight-binding model. This lack of information  is usually not an issue for a single ring resonator since the experiments usually collect signal from a single reference point along the ring.  However, for more complicated structures, e.g. multiple coupled rings, the modal amplitudes are defined with respect to different reference points in different rings, and it would be useful to explicitly describe the connection of these different modal amplitudes. In addition, the tight-binding model in the previous section is a perturbation theory that starts with the modal structure of a unmodulated ring, typically assumed to be lossless. Thus, the model can not accurately describe systems with strong modulation or loss. For these reasons, it is useful to develop a more sophisticated wave-equation approach where the variations of the field amplitude inside the ring are explicitly exhibited. In addition to treating the more complicated system as discussed above, such an approach also allows us to derive the parameters for the tight-binding model in the regime where the tight-binding model is applicable.

To illustrate the wave-equation approach, we provide a more detailed description of the ring under dynamic modulation
at the position $z_{d}$, coupled to external waveguide at the position $z_c$, as shown in Fig. \ref{figure.2}(c). The electric field circulating inside the ring can be expanded as \cite{hausbook}:
\begin{equation}
E (t,r_\perp,z) = \sum_m \mathcal{E}_m (t,z) \mathrm{E}_m (r_\perp) e^{i\omega_m t}, \label{Eq2:E}
\end{equation}
where $z$ denotes the azimuthal position along the ring (such that $\mathcal{E}_m(t,z+L) = \mathcal{E}_m(t,z)$),  $r_\perp$ is the direction perpendicular to $z$, $\mathrm{E}_m (r_\perp)$ is the modal profile of the waveguide, and $\mathcal{E}_m(t,z)$ is the corresponding modal amplitude at the carrier frequency $\omega_m$. Starting from the Maxwell's equations and using  the slowly-varying envelope approximation, one has the wave equation  for the modal amplitude \cite{hausbook}:
\begin{equation}
\left(\frac{\partial}{\partial z} + i\beta(\omega_m)\right)\mathcal{E}_m(t,z) - \frac{1}{v_g}\frac{\partial}{\partial t}{\mathcal E}_m(t,z) = 0, \label{Eq2:wave}
\end{equation}
where $\beta$ is the wavevector of the waveguide. 

The equation for the field at the coupler at the position $z_c$ between a ring and an external waveguide (or another ring) is \cite{fan03}
\begin{equation}
\mathcal{E}_m(t^+,z_c) = \sqrt{1-\gamma_c^2} \mathcal{E}_m(t^-,z_c) - i\gamma_c \mathcal{E}_m^{\mathrm{in}} (t^-,z_c), \label{Eq2:couple}
\end{equation}
where $\mathcal{E}_m^{\mathrm{in}}$ is the amplitude of the input field component at $\omega_m$ from the external waveguide (or another ring), $\gamma_c$ is the coupling strength, and $t^\pm = t+0^\pm$. On the other hand, the equation that describes the field undergoing the dynamic modulation in Eq. (\ref{Eq1:trans}) at the position $z_d$ is \cite{salehbook}:
\begin{equation}
\mathcal{E}_m(t^+,z_d) = \sum_{q=-\infty}^\infty i^q \mathrm{J}_q (2\kappa) \mathcal{E}_{m+q}(t^-,z_d) e^{-iq \Delta t^-}, \label{Eq2:dynmod}
\end{equation}
where $\mathrm{J}_q$ is the Bessel function of the $q$-th order.

The set of equations (\ref{Eq2:wave})-(\ref{Eq2:dynmod}) can be solved numerically in both space and time to provide a more detailed description of the dynamically-modulated ring system \cite{yuan16,yuan18prb,yuan16optica}. Alternatively, to describe the steady state of the system, Eq. (\ref{Eq2:couple}) can be rewritten as a scattering matrix with two inputs and two outputs:
\begin{equation}
S(\gamma_c) = \left( {\begin{array}{*{20}c}
   \sqrt{1-\gamma_c^2} & - i\gamma_c \\
   - i\gamma_c & \sqrt{1-\gamma_c^2} \\
\end{array}} \right). \label{Eq2:Smatrix}
\end{equation}
The wave equation in Eq. (\ref{Eq2:wave}) contributes to a propagation phase for the mode $\mathcal{E}_m$ between two positions $z_1$ and $z_2$ inside the ring:
\begin{equation}
\mathcal{E}_m (z_2) = e^{-i m (z_2-z_1)/v_g} \mathcal{E}_m (z_1).\label{Eq2:propphase}
\end{equation}
Therefore, a discrete linear model from Eqs. (\ref{Eq2:dynmod})-(\ref{Eq2:propphase}) can be constructed to explore the steady state for the system  in Fig. \ref{figure.2}(c) \cite{lin16}.

\section{Experimental methods}

\subsection{Construction of synthetic frequency lattices}

\begin{figure}[htbp]
 \centering
 \includegraphics[width=0.48\textwidth ]{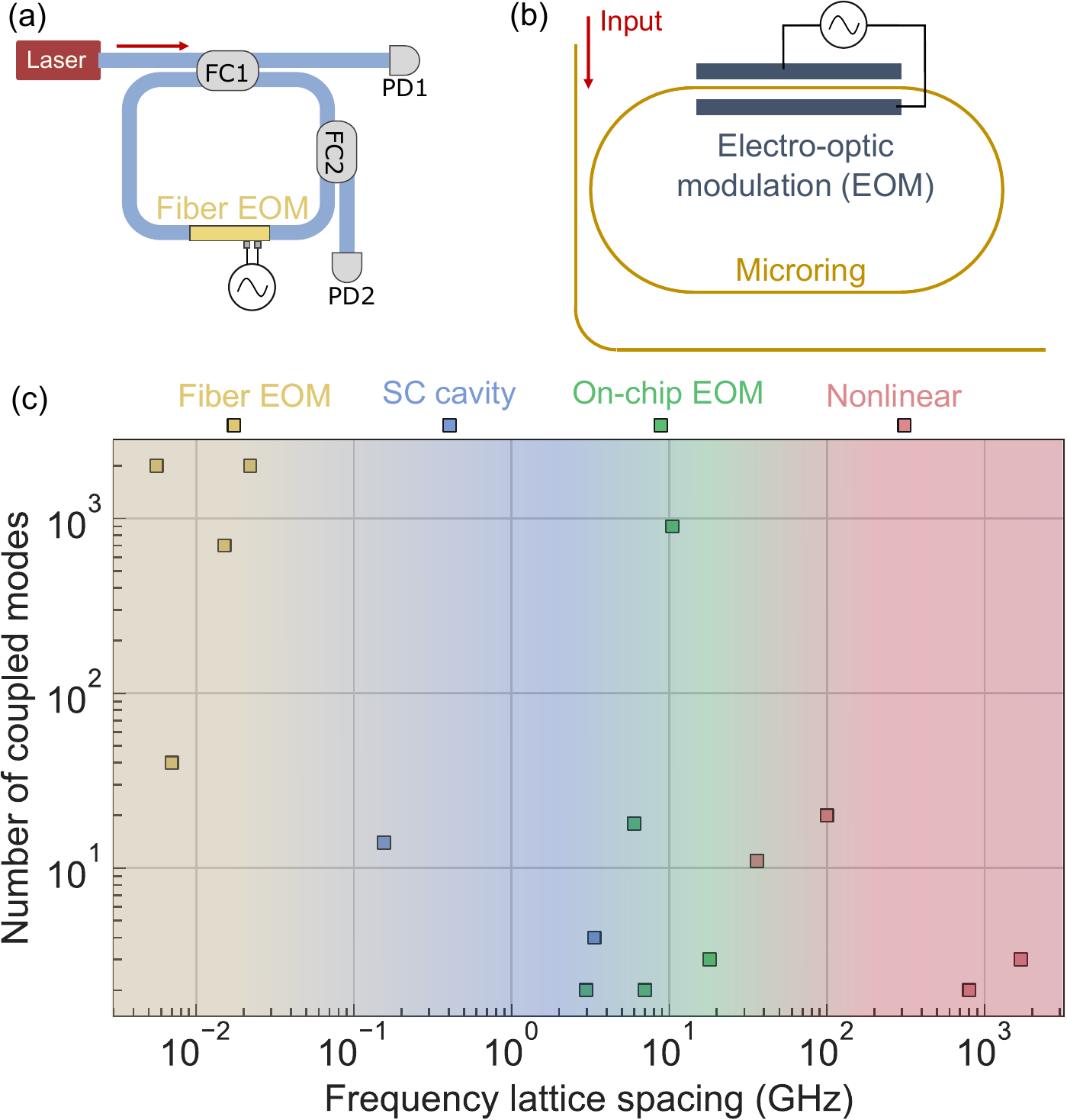}
 \caption{(a) Simplified experimental schematic of a fiber ring resonator with an electro-optic modulator (EOM). Adapted from Ref.~\onlinecite{dutt19NC}. FC1, FC2: fiber splitters/couplers. PD1/PD2: photodetectors. (b) Schematic of an on-chip electro-optically modulated resonator consisting of a waveguide-coupled race-track resonator with electrodes place around it. Modulation is induced by applying an electronic radio-frequency signal. Adapted from Ref.~\onlinecite{hu20}. (c) Frequency lattice spacings and approximate number of coupled modes for experimental demonstrations of frequency synthetic lattices till date. All demonstrations except for the SC cavity (superconducting cavity) are around a center wavelength of 1550 nm. Parameters estimated from reported results in Refs.~\onlinecite{kai21science}, \onlinecite{chen21lsa}, \onlinecite{dutt19NC}, \onlinecite{li21}, \onlinecite{lee20pra}, \onlinecite{tian21arxiv}, \onlinecite{hung_quantum_2021}, \onlinecite{qin18prl}, \onlinecite{zhang19NP}, \onlinecite{hu20}, \onlinecite{lu18prl}, \onlinecite{bersch09}, \onlinecite{bell17}, \onlinecite{joshi20prl}, \onlinecite{joshi18NC} (from left to right, in increasing order of frequency spacing).}
 \label{figure.3}
\end{figure}

The approach as discussed above, where a frequency synthetic dimension is created with the use of dynamically modulated ring resonator structures, can be implemented experimentally in a number of platforms. One platform is based on a fiber resonator incorporating EOMs~\cite{dutt19ACS}, as demonstrated first using commercially available fiber-pigtailed lithium niobate modulators (Fig.~\ref{figure.3}(a))~\cite{dutt19NC} embedded in a ring cavity. In such setups, external fibers are connected to the ring through directional couplers to inject the input laser and readout the cavity field [see Fig. \ref{figure.3}(a)]. In this system the frequency of the input laser can straighforwardly tuned, which  allows for  a controllable detuning $\Delta\omega = \omega_{\rm in} - \omega_0$ of the input laser from the resonator's frequency modes, allowing one to continuously tune from on-resonance excitation to far-off-resonance excitation, in turn enabling direct band structure measurement [see Sec.~\ref{sec:bs}]. Extensions of this simple setup have shown significant potential towards exploring various analogs of condensed matter physics effects that are usually observed for charged electrons but are challenging to observe for neutral photons. For example, it has been demonstrated that one can couple clockwise modes and counter-clockwise modes through additional fibers to realize an effective magnetic flux in two simultaneous synthetic dimensions~\cite{dutt20}. Additionally, one can choose a modulation frequency that is off resonance with the free-spectral range of the ring (i.e. $\Omega_{\rm mod} = \Omega_R + \Delta$, with $\Delta\ne 0$) to introduce an effective electric field~\cite{li21}, or add an amplitude modulator inside the ring to introduce non-Hermitian coupling or to design active mode-locked pulses~\cite{kai21science, yuan18APLP}. This platform has a lattice frequency spacing typically below a few tens of MHz [Fig.~\ref{figure.3}(c)].

As an alternative platform, one can consider a modulated nanophotonic resonator, which holds promise for miniaturization and on-chip integration~\cite{zhang19NP}. 
Towards this end, Hu \textit{et al.}~\cite{hu20} have reported synthetic optical frequency lattices in high dimensions (up to four) based on thin-film nanophotonic lithium niobate microrings using incommensurate modulation frequencies [see Fig.~\ref{figure.3}(b)]~\cite{martin17}. The major requirements for achieving such on-chip integration are the simultaneous realization of: (i) low-loss microrings with a high quality factor, and (ii) a modulation bandwidth higher than the free-spectral range. Both these requirements have been met through improvements in material quality and fabrication techniques of thin-film lithium niobate-on-insulator substrates. On the other hand, acousto-optic modulators and piezoelectric actuation also provide possibilities for strong modulation of rings on chip~\cite{tian21arxiv}.  The successful demonstration of synthetic frequency dimensions with state-of-the-art chips provides a promising candidate for further manipulating the photon's spectrum in the quantum regime. In a completely different frequency range, synthetic dimensions for microwave photons have been reported in a modulated superconducting resonator~\cite{lee20pra, hung_quantum_2021} [see Fig.~\ref{figure.3}(c)] at cryogenic temperatures or a chain of modulated RF cavities at room temperature~\cite{peterson19prl}.

We also note that the frequency synthetic dimension can be implemented in a waveguide without forming the ring resonator.  
Here, one considers a waveguide incorporating a traveling wave modulator. The waveguide supports propagating modes across a continuum of frequencies, and the input frequency $\omega_{\rm in}$ combined with a traveling-wave modulation $\Omega$ defines the frequency grid $\omega_{\rm in} + n\Omega$. In such waveguide systems, the eigenvalues of the Hamiltonian are mapped to the wavevector along the propagation direction of light. While this can simplify the setup compared to the aforementioned resonator implementations where one needs to precisely control the spectral alignment of the input laser with respect to the discrete frequency grid of the ring, the waveguide implementation introduces some limitations in terms of what models can be implemented or excited. For example, since the frequency grid is determined by the excitation itself, it is nontrivial to create a frequency-detuned excitation akin to the resonator implementation. 

The traveling-wave modulation required to create a frequency lattice in waveguides could be realized electro-optically using the Pockels effect (e.g in a lithium niobate waveguide as shown by Qin \textit{et al.}~\cite{qin18prl}) or all-optically using nonlinear four-wave mixing. In fact, the first experiment on frequency lattices used a highly nonlinear fiber driven by a strong pump to study spectral Bloch oscillations of a weak probe field~\cite{bersch09}. Recently, the single pump has been replaced by multiple pumps with controllable relative phases to realize long-range hopping with complex coupling coefficients along the synthetic frequency dimension~\cite{bell17}. All-optical nonlinearity permits very large frequency spacings in excess of 100 GHz to 1 THz [Fig.~\ref{figure.3}(c)].
Wang \textit{et al.} have further developed this platform to construct multidimensional chiral lattices by judiciously designing the pump configuration together with the fiber nonlinearity, to induce nontrivial effective gauge fields \cite{wang20}.  In other experiments, a one-dimensional waveguide array with inhomogeneous couplings between waveguides has synthesized  arbitrary multi-dimensional excitation dynamics, greatly increasing the network dimensionality \cite{maczewsky20}.  

\subsection{Band structure measurement} \label{sec:bs}

With the synthetic frequency dimension in a modulated ring resonator, we have shown that the wavevector along the frequency axis is a time variable. Therefore, the band structure of the synthetic lattice can be quite straightforwardly measured in the ring system coupled to external waveguides in Fig.~\ref{figure.3}(a), by scanning the frequency of the input laser and then measuring the transmitted signals as a function of time~\cite{dutt19NC}. As an example, in case where the ring is modulated with a single near-resonant modulation frequency, using the input-output formalism, one finds that the corresponding transmission of the optical signal is
\begin{equation}
T_{\mathrm{out}}(t=k_f;\Delta\omega) = \frac{\gamma^2}{\left[\Delta\omega-2g \cos(\Omega_R t - \Delta t -\phi)\right]^2+\gamma^2}, \label{Eq4:trans}
\end{equation}
where $\Delta \omega \equiv \omega_{\rm in} - \omega_0$ is the frequency offset between the input frequency $\omega$ and the reference frequency $\omega_0$. For resonant modulation, i.e., $\Delta=0$ or $\Omega = \Omega_R$, Eq. \eqref{Eq4:trans} indicates that the static cosine-like band structure, which is associated with the one-dimensional discrete photonic lattice along the synthetic frequency dimension, can be observed while one performs a time-dependent transmission measurement while scanning the input laser frequency.

\section{Examples of physics studied with synthetic dimensions}

The synthetic dimension concept in modulated ring resonators allows us to synthesize and control a wide range of Hamiltonians, and therefore enables the exploration of a variety of novel physics effects. Here we discuss a few examples of novel physics studied theoretically and experimentally in photonics using synthetic frequency dimensions.

\subsection{Topological photonics in synthetic dimensions}

Topological photonics has become an active field over the past decade, not only to create photonic analogues of nontrivial topological phases that were first explored in condensed matter physics, but also to guide and manipulate light in a manner that is robust to imperfections and disorder~\cite{ozawa19, kim20, leykam20, segev21}. In almost all lattice-based models, the strengths and phases of couplings between lattice sites need to be tailored in a specific fashion to realize topologically nontrivial phases. Additionally, topological phase transitions require the tuning of these couplings to go from a trivial to a nontrivial topological phase.

\begin{figure}[htbp]
 \centering
 \includegraphics[width=0.48\textwidth ]{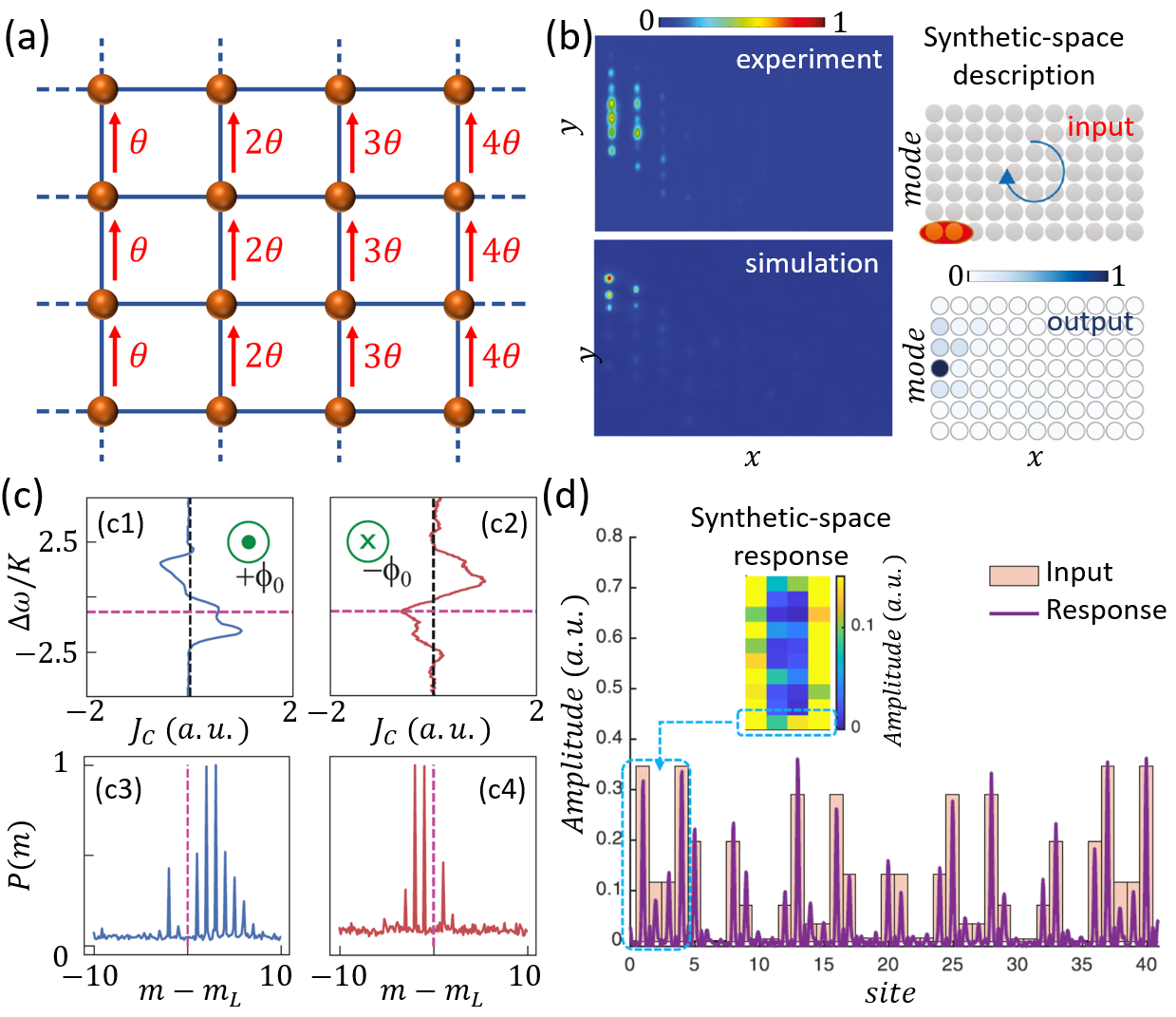}
 \caption{(a) A quantum Hall lattice with a uniform effective magnetic field. (b) Experimental demonstration of quantum Hall physics in waveguide arrays. The input beam excites the boundary sites in synthetic space, which propagates one-way along the synthetic modal dimension at the left boundary. Output fields are collected in experiments, which show excellent agreement with simulations. Adapted from Ref.~\onlinecite{lustig19}.  (c) Measurements of chiral currents in synthetic space in a single ring resonator. (c1) and (c2) Chiral currents $j_C\equiv\sum_{m>m_L} P_{\mathrm{CW}} (m) - \sum_{m<m_L} P_{\mathrm{CCW}} (m)$ versus the laser detuning $\Delta\omega/K$ with different directions of the effective magnetic field. Here $P_{\mathrm{CW(CCW)}}$ is the steady-state photon number of clockwise- (counter-clockwise-) propagating modes at the frequency $\omega_m$, $m_L$ is the order of the resonance closet to the input laser, and $K$ is the strength of the coupling between CW and CCW modes. (c3) and (c4) $P(m)$ for CW and CCW modes at frequencies labelled in dashed lines in (c1) and (c2), showing occupation of predominantly higher and lower frequency modes respectively. Adapted from Ref.~\onlinecite{dutt20}. (d) Measurement of the steady-state edge state in time-multiplexed resonator networks, where a 1D pulse train can be mapped to two synthetic time dimensions. An effective magnetic field with $\theta=2\pi/3$ is applied. Localization of the steady state on the edge of the lattice is observed. Adapted from Ref.~\onlinecite{leefmans21}.}\label{figure.4}
\end{figure}

Synthetic dimensions provide a straightforward way to achieve nontrivial photonic topological phases due to the fact that couplings between discrete modes can be designed nearly at will. As an illustration of how to construct a prototypical topological model -- the quantum Hall lattice, consider a 1D array of modulated ring resonators as shown in Fig.~\ref{figure.4}(a). The discrete resonant modes at frequencies $\omega_m$ in each ring are coupled by the modulation, while the modes at equal frequencies in different rings are coupled evanescently, forming a 2D space with one spatial dimension ($x$) and one frequency dimension ($y$). Moreover, the modulation phase of the modulators are chosen as  $\phi_n= n\theta$ which is linearly dependent on the spatial index of the ring $n$. The distribution of modulation phases is imprinted into hopping phases along the frequency axis of light. Therefore,  a photon circulating around any single square plaquette in the synthetic space accumulates a non-zero phase $\theta$, which corresponds to a uniform effective magnetic flux perpendicular to the plane of the 2D lattice \cite{fang12} (see Fig.~\ref{figure.4}).  This synthetic quantum Hall model supports topologically-protected one-way edge states, which not only facilitate spatial guiding of light without backscattering, but also enable unidirectional frequency conversion~\cite{yuan16,ozawa16}. Similar quantum Hall physics with an effective magnetic flux in synthetic space has also been proposed in a 1D cavity array with the orbital angular momentum (OAM) of light being the synthetic dimension~\cite{luo15} and in a single degenerate cavity with both the frequency and the orbital angular momentum of light being two synthetic dimensions~\cite{yuan19}. Note that the modes along the frequency or OAM dimensions typically lack a well-defined boundary, and hence to observe one-way edge propagation, strong coupling with another resonator~\cite{zhou17,yuan19} or the dispersion engineering of the waveguide forming the ring~\cite{yuan16,shan20} have been proposed to create an artificial boundary.

The first experimental demonstration of quantum Hall physics in photonic synthetic space was reported in a 2D waveguide array, with the synthetic dimension formed by the transverse spatial supermodes of the array. 
In this implementation, the effective magnetic flux was realized by engineering the phases of the hopping coefficients between adjacent supermodes. Topologically-protected one-way edge states in the synthetic space result in bulk propagation in these waveguide arrays [Fig.~\ref{figure.4}(b)] \cite{lustig19}. A quantum Hall ladder has also been demonstrated in a single modulated ring resonator by coupling the clockwise (CW) and counterclockwise (CCW) propagating modes to form a synthetic pseudospin dimension in addition to the frequency dimension. Careful incorporation of auxiliary waveguides with different lengths to couple the CW and CCW spins can  realize an effective magnetic field in this 2D frequency-pseudospin synthetic space. The corresponding chiral current, which is a signature of topological one-way propagation, has been observed in this system, and results in asymmetric frequency conversion (either towards modes with frequencies higher or lower than the input laser frequency) [Fig.~\ref{figure.4}(c)] \cite{dutt20}. As a widely different platform but still built from optical fiber loops, time-multiplexed resonator networks supporting synthetic time dimensions encoded in short pulses \cite{wimmer13,regensburger13,wimmer15,wimmer17,vatnik17,wimmer18,weidemann20} provides another platform to explore effective gauge fields and topological photonics~\cite{chalabi19,chalabi20,leefmans21}. The complex connectivity between pulses with different arrival times can be controlled by modulators in additional fiber loops. Moreover, long-range couplings can be introduced by proper choice of additional waveguides, so the 1D train of pulses can be mapped to a 2D synthetic space. A topological response in such a 2D lattice has been measured [Fig.~\ref{figure.4}(d) ]~\cite{leefmans21}, with significant prospects for extensions to much higher-dimensional lattices.

Besides the quantum Hall effect with an effective magnetic flux in 2D square lattice, various topological phases with different dimensionalities have been explored using photonic synthetic dimensions. A Haldane model has been studied purely in synthetic space using the frequency axis of light~\cite{yuan18prb}. Graphene dynamics can be simulated in a 1D array of rings with staggered resonances, showing potential capability for constructing synthetic 2D lattice with C$_3$ symmetry \cite{yu21graphene}. Beyond 2D lattices, 3D topological phases such as a photonic Weyl point~\cite{lin16,zhang17,sun17} and a topological insulator~\cite{lin18} have been proposed in a 2D array of resonators together with a synthetic dimension. Moreover, the four-dimensional quantum Hall effect was proposed in a 3D array of resonators with a frequency dimension~\cite{ozawa16}. The 4D quantum Hall effect is an example of topology with no-lower dimensional analogue, and it highlights the advantage of synthetic space implementations to realize phases beyond what is possible in the three real spatial dimensions. Recently, synthetic dimensions have been harnessed to explore higher-order topology \cite{dutt20light,zhang20} and topological non-equilibrium quench dynamics \cite{yu21arxiv}.

\subsection{Band structure in the synthetic dimension}  \label{sec:bsexpt}

\begin{figure}[htbp]
 \centering
 \includegraphics[width=0.48\textwidth ]{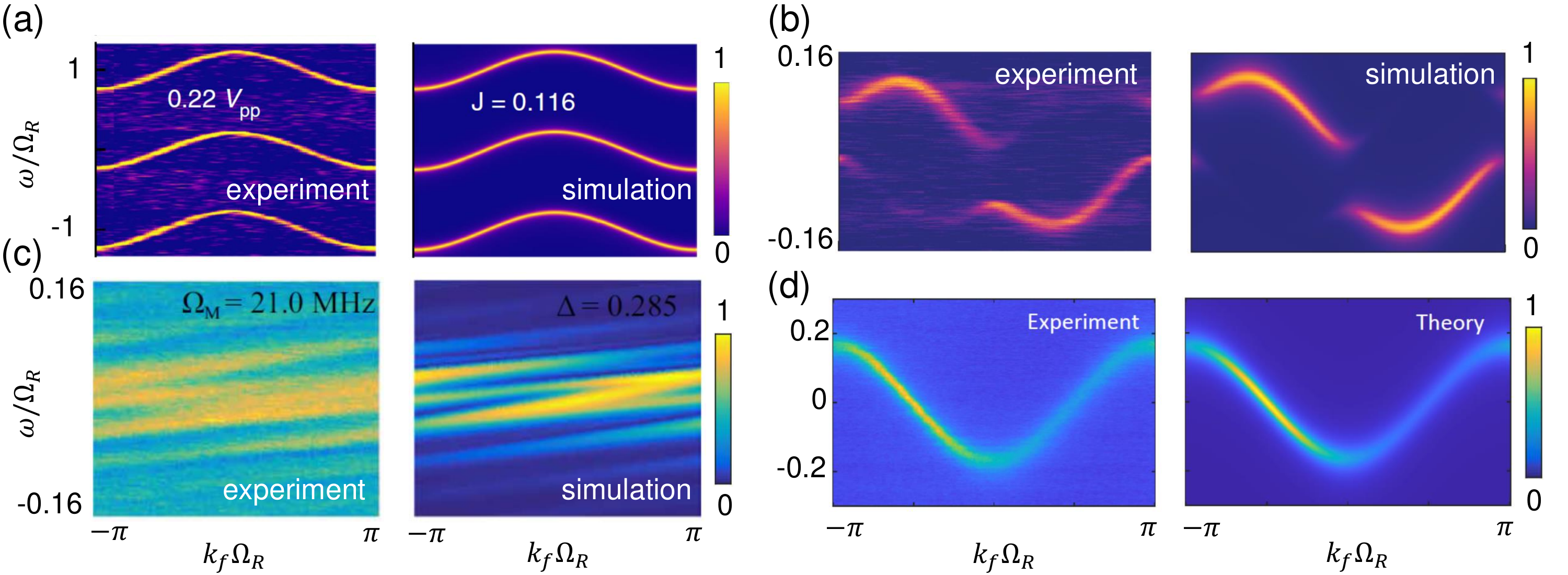}
 \caption{Band stucture measurement in the synthetic dimension. (a) Direct measurement of band structure in a synthetic lattice along the frequency axis of light. 
 (b) Measurement of topological edge mode in the synthetic space. 
 (c) Measurement of a dynamic band evolution in the synthetic frequency dimension for a modulation detuned from the FSR by $\Delta = \Omega_R - \Omega_{M} = 2\pi \times$1 MHz.  
 (d) Measurement of non-Hermitian band structure. 
 Panels (a), (b), (c) and (d) are adapted from Refs.~\onlinecite{dutt19NC}, \onlinecite{dutt20}, \onlinecite{li21} and \onlinecite{kai21science} respectively.}\label{figure.5}
\end{figure}

As we have shown in Section~\ref{sec:bs}, the dynamically modulated ring resonator system provides a straightforward method to measure the band structure in the synthetic frequency dimension. Here we show several examples. Fig.~\ref{figure.5}(a) shows the measured cosine-like band structure of a 1D tight-binding lattice with nearest-neighbor coupling, implemented in a frequency synthetic dimension using relatively simple optical fiber technology~\cite{dutt19NC}. This was in fact the first demonstration of a direct band structure measurement in synthetic space, which matches very well with the prediction by Eq.~\eqref{Eq4:trans}. In this experiment, long-range coupling with a modulation frequency $2\Omega$ was also incorporated, with a modulation phase $\theta$. This results in complex-valued coupling between the next-nearest-neighbor modes along the frequency dimension with a phase $\theta$, creating a nonreciprocal band structure that breaks time-reversal symmetry. The technique of reading out the band structure from a time-resolved transmission measurement has found several extensions to more complicated models, including those with two bands, effective electric forces, and non-Hermitian coupling. 

For a model with two bands, such as a quantum Hall ladder formed using two independent synthetic dimensions of frequency and spin in a single ring, the time-resolved transmission gives information of the projected band structure to the left or the right leg of the Hall ladder [see Fig.~\ref{figure.5}(b)]~\cite{dutt20}. The band-structure measurement technique clearly reveals spin-momentum locking and the topological chiral nature of the bands [Fig.~\ref{figure.5}(b)]. On the other hand, for a non-resonant modulation, i.e., a non-zero $\Delta$ in Eq.~\eqref{Eq4:trans}, an effective electric force along the frequency dimension is created. Interesting physical implications of this force are the Bloch oscillations in lattice space (which is the frequency here) \cite{yuan16optica,bersch09,hu20,longhi05,peschel08,bersch11}, and a dynamic band evolution in reciprocal space. Such a dynamic band structure evolution was recently measured in the  modulated ring system as shown in Fig. \ref{figure.5}(c) \cite{li21}.

For the lossless (i.e. Hermitian) Hamiltonians that we considered so far, the coupling coefficient $\gamma$ between the ring and the external waveguide broadens the transmission peak by a $k$-independent constant width $2\gamma$. This technique of band structure readout from a time-resolved transmission can also be generalized to non-Hermitian bands, where $\gamma$ is now replaced by a $k$-dependent imaginary part of the band energy $\mathrm{Im}(k)$ in Eq.~\eqref{Eq4:trans}. A Lorentzian fit at each $k$ then reveals both the real part and the imaginary part of a non-Hermitian band [Fig.~\ref{figure.5}(d)]~\cite{kai21science}. Hence, the synthetic-space band-structure measurement technique opens a pathway for understanding as well as synthesizing the band structure of a periodic system.

\subsection{Towards other novel physics and applications in synthetic dimensions: non-Hermitian, nonlinear and quantum photonics}

Non-Hermitian systems are natural in photonics due to the ubiquity of loss and interactions of photonic modes with the environment they are embedded in. Early work in non-Hermitian photonics focused on parity-time symmetry, with the possible inclusion of gain and active media to induce intriguing lasing properties. Specifically in the context of topology, recent work has revealed unique kinds of topology found exclusively in non-Hermitian systems due to the complex-valued eigenenergies that characterize non-Hermitian bands. For example, complex-valued band energies allow for topological invariants defined in terms of the energy \textit{eigenvalues}, instead of the usual Hermitian topological invariants defined in terms of energy \textit{eigenstates}~\cite{bergholtz21}. Such non-Hermitian lattice models have been constructed in photonics in real or synthetic space, as well as in acoustics and electrical circuits, but remain challenging to realize in condensed matter systems.

Synthetic dimensions provide a fertile ground to explore non-Hermitian physics, and have led to theoretical proposals for engineering active mode locking~\cite{longhi16} and pulse shortening~\cite{yuan18APLP}. Using an array of modulated ring resonators with spatially engineered gain, a topological insulator laser has been studied in synthetic frequency dimensions, which shows promise for actively mode-locking many rings to realize robust high-power pulsed lasers~\cite{yang20}. Moreover, the 2D non-Hermitian skin effect has also been examined in a synthetic photonic lattice. In this model, second-order corner modes have been shown, which can be used to manipulate light in both spatial and spectral domains~\cite{song20}. 

On the experimental side, as briefly mentioned in Sec.~\ref{sec:bsexpt}, non-Hermitian bands in a frequency synthetic dimension have been characterized to observe arbitrary topological winding of the band in the complex plane \cite{kai21science}, and hence to characterize topologically nontrivial phases in non-Hermitian systems. A parity-time symmetric photonic system in  wavelength space has been shown in the microwave regime, which leads to single-mode oscillation of the optoelectronic oscillator in a simple experimental configuration \cite{zhang20NC}.

A prime goal of photonic lattices is the simulation of many-body interacting phases of light and matter. 
This is significantly challenging due to the weakly interacting nature of photons, and strong nonlinearities mediated by novel material platforms are hence necessary to overcome this. An additional problem that is particular to synthetic dimension implementations such as frequency or OAM, is the long-range nature of the interaction, since all the modes along the synthetic dimension are co-located within the same waveguide or resonator.
A method was proposed recently to address these issues by introducing strong group-velocity dispersion into a highly nonlinear ring resonator.
Such a method achieves a Hamiltonian where the interactions are completely local and allows the photon blockade effect to be theoretically studied in the synthetic frequency space~\cite{yuan20}, paving the way for quantum many-body physics in future studies. In the classical regime, intriguing states such as chimera states and dynamical solitons in frequency combs have been proposed by incorporating third-order nonlinearities into a modulated ring resonator~\cite{tusnin20}.

Other novel physics has also been widely explored in synthetic dimensions. 1D Bloch oscillation along the synthetic frequency dimension has been studied \cite{yuan16optica, longhi05}, which can be further generalized towards a technique for generating broadband frequency-modulated light \cite{harris20}. While steady-state signatures of Bloch oscillations have been reported previously in nonlinear fibers~\cite{bersch09}, such oscillations have recently been observed in real-time using modulated ring resonators~\cite{chen21lsa}. On the other hand, coherent random walks of photons have been experimentally realized in different platforms such as the time-multiplexed system~\cite{chen18prl}, the synthetic frequency dimension~\cite{hu20}, and the OAM dimension~\cite{wang18prl}.

Synthetic dimensions are promising for manipulating genuine quantum interference and entanglement between photons along internal degrees of freedom. Towards this end, efficient manipulation of high-dimensional quantum entanglement has been proposed in synthetic space by combining frequency and OAM dimensions~\cite{yuan19}. Moreover, unitary transformations for photons along the synthetic frequency dimension has been investigated using electro-optic modulation or with time-resolved photon detection, with prospects for frequency-encoded quantum information processing (QIP)~\cite{buddhiraju21NC,lu2019ieee,cui2020prl}. On the experimental front, rapid progress in frequency-encoded QIP has been reported using both electro-optic modulation as well as nonlinear wave mixing~\cite{pysher11PRL,roslund14NP,kues17nature,lu18prl,joshi18NC,joshi20prl}, but several challenges regarding loss and scalability remain.
As opposed to frequency-encoding, an alternative platform for scalable, deterministic quantum computation has been proposed using a synthetic time dimension in a ring loop coupled to a single cavity-coupled atom or quantum emitter~\cite{bartlett21arxiv}. A very similar loop-based time-multiplexed architecture also finds potential in creating a photonic neural network~\cite{peng21arxiv}. Furthermore, time-multiplexing has resulted in entanglement generation between more than a million quantum modes, which is the largest number of entangled modes till date~\cite{yoshikawa16apl}.
These recent developments indicate that the concept of synthetic dimensions, which was inspired by fundamental studies in condensed matter topology and ultracold atoms, may lead to  emerging practical applications of photonics, with significant technological implications for both classical and quantum technologies.


\section{Summary and outlook}

In this tutorial, we overview several theoretical approaches and experimental platforms for realizing synthetic dimensions in photonics by connecting different resonant modes in the dynamically modulated ring resonator system. Recent examples of physics studied with synthetic dimensions have also been reviewed. Research associated with the creation of synthetic dimensions with complex connectivity in photonics has witnessed rapid growth in recent years. The idea for connecting discrete optical modes to form a synthetic photonic lattice have several attractive aspects: First, synthetic dimensions provide a way to explore high-dimensional physics, such as the four-dimensional Hall effect \cite{ozawa16}, in a lower-dimensional physical structure. Moreover, interesting physics can be explored in a relatively simple experimental platform, which attracts great significance in quantum simulations. Second, the connectivity between discrete modes can be readily designed to achieve complicated functionalities, such as the effective magnetic field for photons \cite{yuan16,ozawa16,lustig19,dutt20} and complex long-range coupling \cite{bell17, yuan18prb, dutt19NC}, which are otherwise difficult to achieve in spatial dimensions. Lastly, physical phenomena in synthetic dimensions can be used to manipulate photons in novel ways. For example, the one-way edge states enable unidirectional frequency conversion that is topologically protected \cite{yuan16}. Along this line, the properties of photons, including the frequency, orbital angular momentum, and arrival time of pulses, can be controlled using a very different perspective, which is promising for applications in optical communications and optical information processing. 
%
%
Therefore, besides seeking for more opportunities in studying new physics and manipulating photons, future investigations into synthetic dimensions would benefit from a push towards the quantum regime \cite{cheng21arxiv}, for quantum computation as well as miniaturization of photonic devices towards on-chip applications.



\begin{acknowledgments}
This work is supported by National Natural Science Foundation of China (11974245), Natural Science Foundation of Shanghai (19ZR1475700), a Vannevar Bush Faculty Fellowship (grant N00014-17-1-3030) from the U.S. Department of Defense, and MURI grants from the U.S. Air Force Office of Scientific Research (grants FA9550-17-1-0002 and FA9550-18-1-0379). L.Y. acknowledges support from the Program for Professor of Special Appointment (Eastern Scholar) at Shanghai Institutions of
Higher Learning.
\end{acknowledgments}

\end{document}